\documentclass[10pt,a4paper]{book}
\usepackage{udineHyderabad-Klu2}

\begin{document}

	\pagestyle{fancy}
	
\talktitle{Dark matter detection in gamma astroparticle experiments}{Dark matter detection in gamma astroparticle experiments}

\talkauthors{Erica Bisesi\structure{a},
	Mosè Mariotti\structure{b},\\
	Villi Scalzotto\structure{b}}

\authorstucture[a]{Department of Physics, Udine University and INFN, Italy}
		   
\authorstucture[b]{Department of Physics, Padua University and INFN, Italy}

\shorttitle{Dark matter detection in gamma astroparticle experiments} 

\firstauthor{E.\ Bisesi et al.}

		\index{Bisesi@\textsc{Bisesi}, E.}
		\index{Mariotti@\textsc{Mariotti}, M.}
		\index{Scalzotto@\textsc{Scalzotto}, V.}

\begin{abstract}
The content of matter in the Universe is estimated to be the 27\% of its 
critical density. 
It is almost universally accepted that most ot this matter is non-baryonic. 
Constraints from primordial nucleosynthesis and cosmic background radiation 
measurements impose that the baryonic content of the Universe cannot exceed 
the 4\% of the critical density, so the nature of the remaining 23\% has yet 
to be identified. In this sense, one of the most promising candidates is 
represented by supersymmetric neutralinos. If they exist, they give rise to 
relic densities in the required range, and are very well motivated in the 
framework of theoretical extensions of the Standard Model of particle
physics. 
In addition to direct neutralino searches and collider experiments, neutralino 
annihilation into gamma rays, neutrinos and synchrotron emission from the 
charged products represents a reliable way of detecting these intriguing
particles. The strongest signals are expected to come from the Galactic
Center and from the nearest dwarf spheroidals. Clumps of dark matter in
galactic haloes are well predicted by high resolution cold dark matter 
numerical simulations. In this work we present our studies on the gamma-ray 
emission from the Galactic Center and from the Draco dwarf spheroidal. We 
investigate the effect of clumpiness on the detection of signals from 
neutalinos for different mass density profiles. One of the scientific goals 
of the MAGIC telescope are just searches for the stable lightest
supersymmetric particle in the different physical scenarios in which they are 
produed. Assuming MAGIC specifications, we draw some conclusions about the 
potentialities of this telescope in such a kind of investigation.
\end{abstract}

\section{Introduction}  
One of the most promising candidates for halo dark matter are weakly
interacting massive particles (WIMPs), although it is not excluded that 
particles of other kinds, not yet predicted by particle physics models, might 
represent the solution of this issue. These particles give a relic density 
which is of the right order of magnitude to explain the dark matter on all
scales. Neutralino is the lightest stable supersymmetric particle in most 
models, so we focus on detection prospects of such a candidate, working either 
in the MSSM or in the mSUGRA frameworks.

High resolution numerical simulations of dark haloes formation suggest that
the strongest signals are likely to come from the Galactic Center and from the 
nearest dwarf spheroidals. The persistence of substructures in these
simulations induces to argue that at least a fraction of the dark matter in
haloes is clustered in clumps.

Taking a phenomenological approach, we here discuss the implications
of\linebreak[5] \mbox{clumpiness on neutralino} dark matter searches.

\vfil

\section{Particle physics models}
Minimal supersymmetric model has many free parameters, but with some
assumptions we are left with seven parameters in the MSSM model and with five
parameters in the mSUGRA setup, namely the supersymmetric extension of the
Standard Model defined in a supergravity inspired framework. For details about
the parameters which fully define the action of MSSM and mSUGRA see Ref.~\protect\cite{6bis,7bis}.

Table 1 shows the range of parameter values used in our scan of the MSSM
space. Choosing the cosmologically interesting relic density range $0.094 <
\Omega_{\chi} h^{2} < 0.129$, we generate in this framework 500000 models 
and impose that they are not excluded by accelerators constraints.

We sample the 5-dimensional mSUGRA parameter space choosing a few values of
$tg\,\beta$ and $A_{0}$, and slices along the $m_{1/2}, m_{0}$ plane for both
$sign\,\mu$. We consider both the slepton and the stop coannihilation regions 
and calculate the relic density with all coannihilations. 
Visited benchmark points are indicates at the upper left of Fig.\ (3).

\begin{table}[!hbp]
\centering
\caption{Scans of the MSSm space.}
\begin{tabular}{c c c c c c c c } 
\multicolumn{8}{c}{}\\\hline\hline
\footnotesize{Parameter} & \footnotesize{$\mu$}  & \footnotesize{$M_{2}$} &
\footnotesize{$tg\beta$}  & \footnotesize{$m_{A}$} & \footnotesize{$m_{0}$}  &
\footnotesize{$A_{b}/m_{0}$} & \footnotesize{$A_{t}/m_{0}$}\\
\footnotesize{Unit} & \footnotesize{GeV}  & \footnotesize{GeV} &
\footnotesize{1}  & \footnotesize{GeV} & \footnotesize{GeV}  &
\footnotesize{1} & \footnotesize{1}\\\hline
\footnotesize{Min} & \footnotesize{10}  & \footnotesize{10} &
\footnotesize{1}  & \footnotesize{10} & \footnotesize{50}  &
\footnotesize{-3} & \footnotesize{-3}\\
\footnotesize{Max} & \footnotesize{10000}  & \footnotesize{10000} &
\footnotesize{60}  & \footnotesize{1000} & \footnotesize{10000}  &
\footnotesize{3} & \footnotesize{3}\\\hline\hline
\end{tabular}
\end{table}

\section{Dark matter distribution models}

We focus on indirect searches of neutralinos in the Milky Way and in the Draco
dwarf spheroidal. We model haloes of these galaxies using the following dark 
matter profiles:
\begin{itemize}
\item  \textbf{Navarro$-$Frenk$-$White} cuspy model \protect\cite{4bis}: $\rho_{cusp}(r) = \frac{\rho_{0}}{(r/r_{s})^{\gamma}\,(1+(r/r_{s}))^{3-\gamma}}$, $\gamma=1$;
\item  \textbf{Moore \& al.} cuspy model \protect\cite{3bis}: the same as above with $\gamma=1.5$;
\item a \textbf{milder cuspy} profile \protect\cite{2bis}: the same as above with $\gamma=0.5$;
\item \textbf{Burkert \& al.} profile \protect\cite{5bis}: $\rho_{Burk}(r) = \frac
{\rho_{0}}{(1+(r/r_{s}))\, (1+(r/r_{s})^{2})}.$   
\end{itemize} 

\section{Gamma ray flux}
Neutralino annihilation in the galactic halo produces both a gamma-ray flux
with a continuum energy spectrum and monochromatic gamma-ray
lines. Considering a detector with an angular acceptance $\Delta\Omega$
pointing in a direction of galactic longitude and latitude $(l,b)$, the
gamma-ray flux from neutralino annihilation at a given energy $E$ is:
\begin{equation}
\label{1.1bis}
\Phi_{\gamma}(E,\Delta\Omega,l,b) = const.\, \sum_{F}
\frac{<\sigma\,v_{F}>}{m_{\chi}^{2}}\, \frac{dN_{\gamma}^{F}}{dE}\, <J(l,b)>\,
(\Delta\Omega) \mbox{ cm$^{-2}$\, s$^{-1}$\, sr$^{-1}$}
\end{equation}
for the continuum gamma. If we assume a spherical dark matter halo in the form
of a perfectly smooth distribution of neutralinos, $<J(l,b)>(\Delta\Omega)$ is
equal to:
\begin{equation}
\label{1.2bis}
<J(l,b)>(\Delta\Omega) = const'\, \frac{1}{\Delta\Omega}\,
\int_ {\Delta \Omega} d\Omega'\, \int_{line of sight} \rho(L,\psi')^{2}dL,
\end{equation}
where $L$ is the distance from the detector along the line of sight, $\psi$ is
the angle between the direction of observation and that of the center of the
galaxy and the integration over $d\Omega'$ is performed over the solid
angle $\Delta\Omega$ centered on $\psi$.
 
\subsection{Effect of clumpiness}
To discuss the implications of clumpiness on neutralino dark matter
searches, we follow the phenomenological approach of Bergstr\"{o}m \&
al.\ (1999) \protect\cite{1bis}. This model is mainly focused on a many small clumps 
scenario, where substructures are light, with $M_{cl}$ less than $10^{4}-10^{6}
M_{\odot}$. Postulating that a fraction $f$ of the dark matter is concentrated 
in clumps, they find that the increase of the signal compared to a smooth halo 
is determined by the enhancement factor $f\,\delta$, where $\delta$ is the
effective contrast between the dark matter density in clumps and the local
halo density $\rho_{0}$. Assuming that the clumps can be regarded as pointlike
sources, authors derive the anoalogous of Eq.\ (2) in the clumpy scenario:
\begin{equation}
\label{1.3bis}
<J(l,b)>(\Delta\Omega) = const'\, \frac{f\,\delta}{\Delta\Omega}\, \rho_{0}\,
\int_ {\Delta \Omega} d\Omega'\, \int_{line of sight} \rho(L,\psi') dL.
\end{equation}   

\begin{figure}[!t]
\centering
\includegraphics[width=6.1cm]{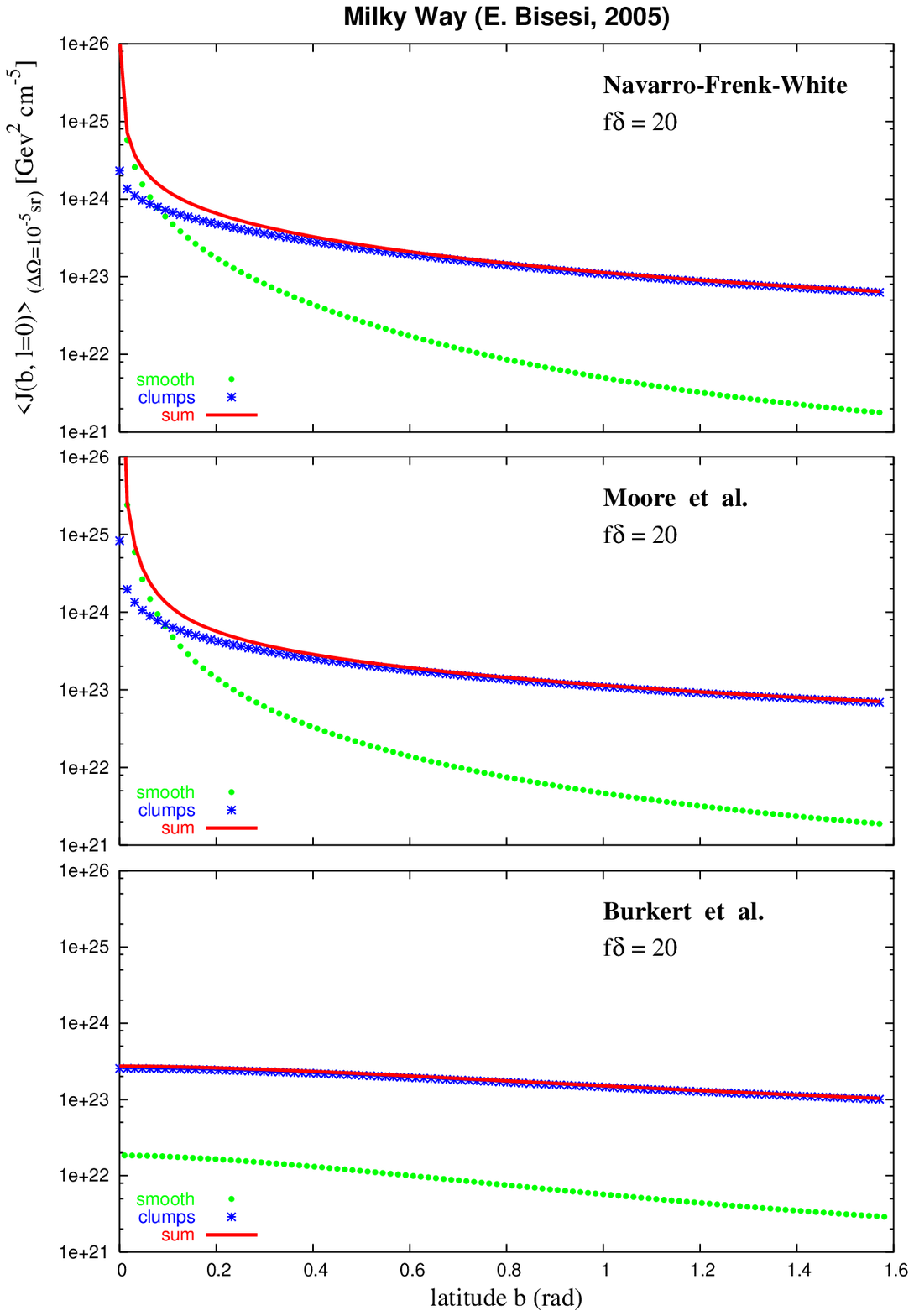}
\includegraphics[width=6.1cm]{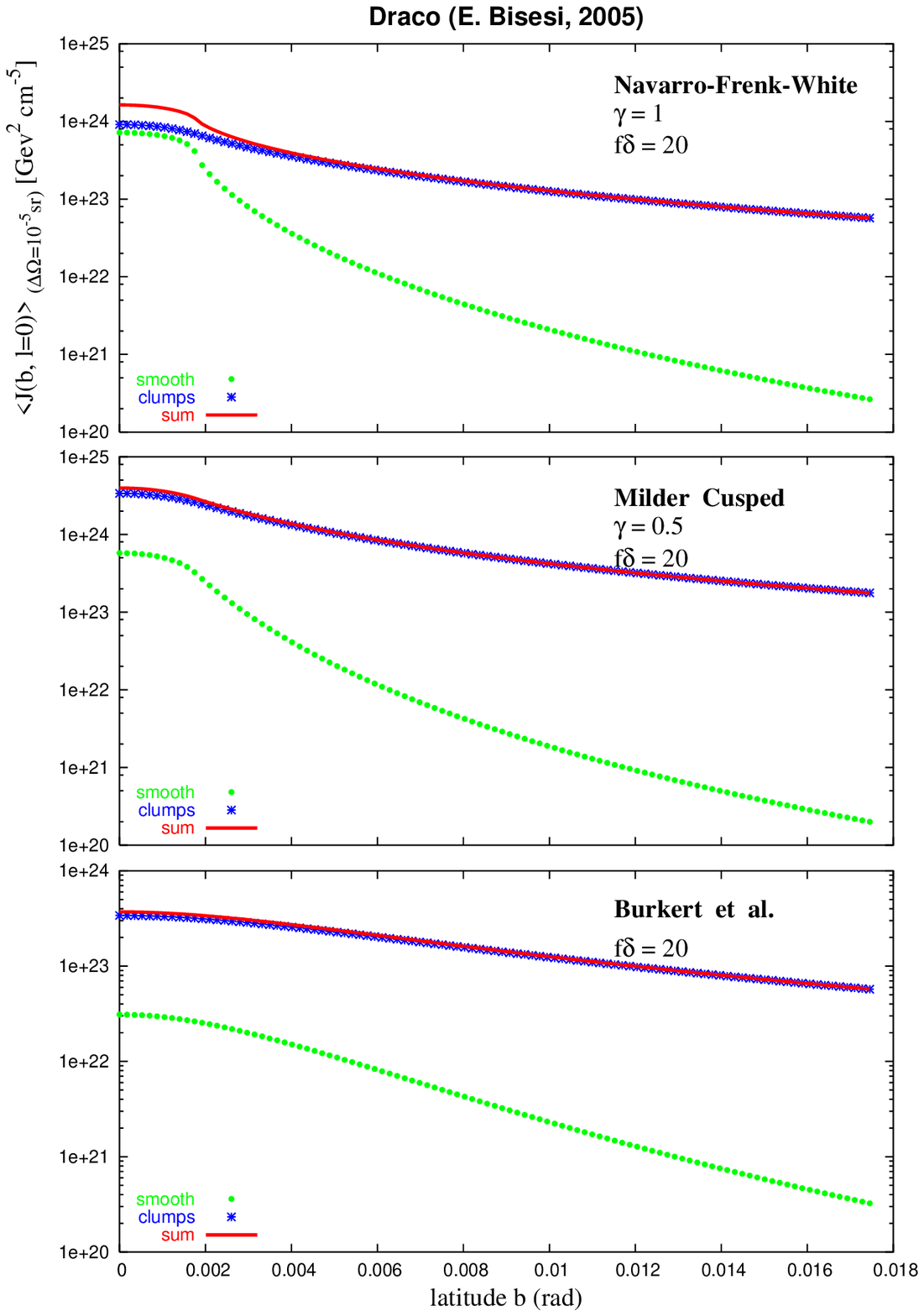}
\caption{ Values of $<J(l,b)>(\Delta\Omega)$ for different halo profiles for
the Milky Way and Draco.}
\label{Fig.1bis}
\end{figure}

Fig.~(\ref{Fig.1bis}) illustrates the values of $<J(l,b)>(\Delta\Omega)$ for 
three of the halo profiles introduced above, giving the smooth and clumpy 
components in the cases of the Milky Way and Draco respectively. The
clumpiness enhancement factor is taken reasonably equal to 20.
Values of the scale lenght and local halo densities follow prescriptions of
 Ref.~\protect\cite{1bis,2bis}.

\section{Results and discussion}
\begin{figure}[!t]
\centering
\includegraphics[width=9.05cm]{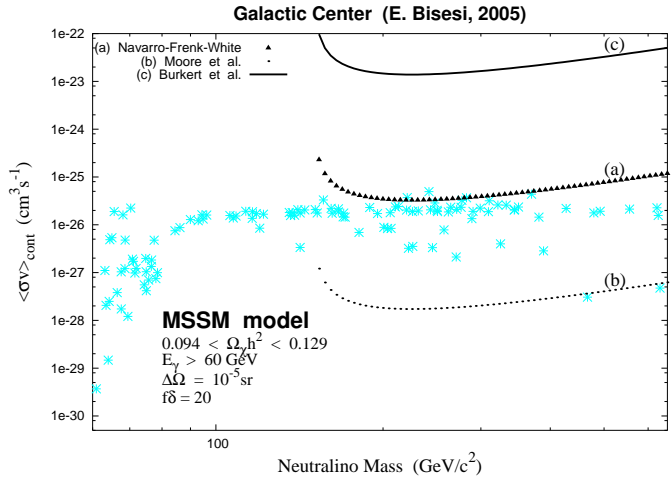}
\caption{ The minimum detectable $<\sigma\,v>_{cont}$ versus $m_{\chi}$ for
the NFW, Moore \& al. and Burkert \& al. profiles. Dots are points of the
parameter space of MSSM, lines represent the $5\sigma$ detection level
for the MAGIC telescope. Only models corresponding to SUSY points above the
curves yield a detectable signal.}
\label{Fig.2bis}
\end{figure}

\begin{figure}
\centering
\includegraphics[width=9.05cm]{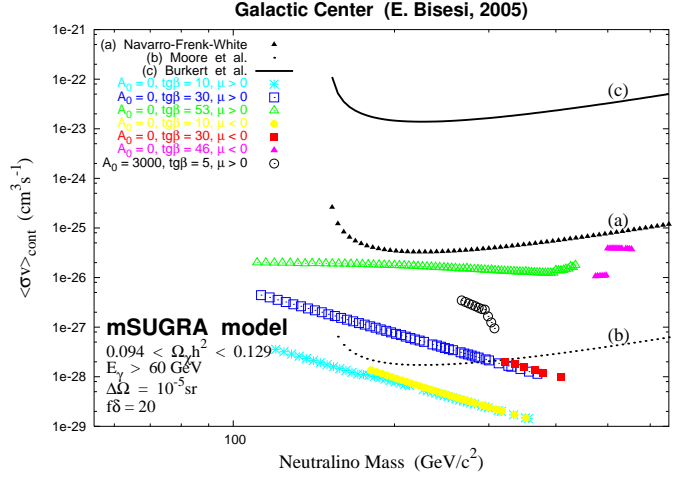}
\caption{ The minimum detectable $<\sigma\,v>_{cont}$ versus $m_{\chi}$ for
the NFW, Moore \& al.\ and Burkert \& al.\ profiles. Dots are points of the
parameter space of mSUGRA, lines represent the $5\sigma$ detection level
for the MAGIC telescope. Only models corresponding to SUSY points above the
curves yield a detectable signal.}
\label{Fig.3bis}
\end{figure}

As the background follows a poissonian statistics, the minimum detectable flux
of gamma rays from an ACT telescope is determined by the condition:
\begin{equation}
\label{1.4bis}
\frac{\Phi_{\gamma}\,A_{eff}\,t\,\Delta\Omega}{\sqrt{N_{b}}} \ge 5,
\end{equation} 
for a 5$\sigma$ detection level. $N_{b}$ is the number of background counts,
hadrons and electrons, which is obtained on the ground of Ref.~\protect\cite{1bis}. 
We assume the following MAGIC specifications: $E_{th}=60$ GeV, 
$A_{eff}=10^{9}$ cm$^{2}$, $t=250$ h, $\Delta\Omega=10^{-5}$ sr and an energy 
resolution of 25\%. Plotting inequality (4) with the equality sign onto the 
SUSY parameter space, we divide it into the detectable (above the line) and 
undetectable (below the line) regions. Results for the Galactic Center are 
shown in Fig.~(\ref{Fig.2bis}) and Fig.~(\ref{Fig.3bis}), for the MSSM and mSUGRA 
scenarios respectively and for a clumpiness enhancement factor of 20. This
factor is anyway uninfluential in this case.

As we can see from Fig.~(\ref{Fig.2bis})$-$(\ref{Fig.3bis}) plots,
detectability of SUSY particles is very sensitive to the choice of the dark
matter profile. We find that the scenario which gives the best opportunities 
for the MAGIC telescope is the Navarro-Frenk-White. Anyway, our model 
doesn't take into account distribution functions for substructures; an
extension of our investigations at higher galactic latitudes does need this is 
to be modeled in detail, so we will address our future interests in this 
direction.

\end{document}